\documentclass[%
 reprint,
superscriptaddress,
 amsmath,amssymb,
 aps,
]{revtex4-1}

\usepackage{graphicx}
\usepackage{dcolumn}
\usepackage{bm}

\begin{document}

\preprint{APS/123-QED}

\title{First Measurement of the Atomic Electric Dipole Moment of $^{225}$Ra}

\author{R.H. Parker}
\affiliation{Physics Division, Argonne National Laboratory, Argonne, Illinois 60439, USA}
\affiliation{Department of Physics and Enrico Fermi Institute, University of Chicago, Chicago, Illinois 60637, USA}
\author{M.R. Dietrich}
\affiliation{Physics Division, Argonne National Laboratory, Argonne, Illinois 60439, USA}
\affiliation{Department of Physics and Astronomy, Northwestern University, Evanston, IL 60208, USA}
\author{M.R. Kalita}
\affiliation{Physics Division, Argonne National Laboratory, Argonne, Illinois 60439, USA}
\affiliation{Department of Physics and Astronomy, University of Kentucky, Lexington 40506, USA}
\author{N.D. Lemke}
\altaffiliation[Present address: ]{Space Dynamics Laboratory, Logan, UT 84341, USA}
\affiliation{Physics Division, Argonne National Laboratory, Argonne, Illinois 60439, USA}
\author{K.G. Bailey}
\affiliation{Physics Division, Argonne National Laboratory, Argonne, Illinois 60439, USA}
\author{M. Bishof}
\affiliation{Physics Division, Argonne National Laboratory, Argonne, Illinois 60439, USA}
\author{J.P. Greene}
\affiliation{Physics Division, Argonne National Laboratory, Argonne, Illinois 60439, USA}
\author{R.J. Holt}
\affiliation{Physics Division, Argonne National Laboratory, Argonne, Illinois 60439, USA}
\author{W. Korsch}
\affiliation{Department of Physics and Astronomy, University of Kentucky, Lexington 40506, USA}
\author{Z.-T. Lu}
\email[Corresponding author: ]{lu@anl.gov}
\affiliation{Physics Division, Argonne National Laboratory, Argonne, Illinois 60439, USA}
\affiliation{Department of Physics and Enrico Fermi Institute, University of Chicago, Chicago, Illinois 60637, USA}
\author{P. Mueller}
\affiliation{Physics Division, Argonne National Laboratory, Argonne, Illinois 60439, USA}
\author{T.P. O\textsc{\char13}Connor}
\affiliation{Physics Division, Argonne National Laboratory, Argonne, Illinois 60439, USA}
\author{J.T. Singh}
\affiliation{Physics Division, Argonne National Laboratory, Argonne, Illinois 60439, USA}
\affiliation{National Superconducting Cyclotron Laboratory and Department of Physics \& Astronomy, Michigan State University, East Lansing, Michigan 48824 USA}

\date{\today}

\begin{abstract}
The radioactive radium-225 ($^{225}$Ra) atom is a favorable case to search for a permanent electric dipole moment (EDM).  Due to its strong nuclear octupole deformation and large atomic mass, $^{225}$Ra is particularly sensitive to interactions in the nuclear medium that violate both time-reversal symmetry and parity. We have developed a cold-atom technique to study the spin precession of $^{225}$Ra atoms held in an optical dipole trap, and demonstrated the principle of this method by completing the first measurement of its atomic EDM, reaching an upper limit of $|$$d$($^{225}$Ra)$|$ $<$ $5.0\!\times\!10^{-22}$ $e \cdot$cm (95$\%$ confidence). 
\begin{description}
\item[PACS numbers]
11.30.Er, 24.80.+y, 32.10.Dk, 37.10.Gh
\end{description}
\end{abstract}

\pacs{Valid PACS appear here}
\maketitle


The existence of a permanent electric dipole moment (EDM) would violate both time-reversal ($T$) and parity ($P$) symmetries \cite{pr50}, and, by the $CPT$ theorem, charge-parity ($CP$) symmetry \cite{CPwoS}. Although the Standard Model (SM) exhibits $CP$ violation \cite{ccft64,babar,belle}, its effects contribute to EDMs only at higher orders \cite{CPwoS}, resulting in EDM values too small to be detected in the foreseeable future. On the other hand, extensions of the SM, such as supersymmetry, naturally provide additional sources of $CP$ violation that explain the matter-antimatter asymmetry of the universe and result in observable EDMs \cite{fsb03}. Limits derived from recent EDM searches in the neutron \cite{baker06}, the $^{199}\mathrm{Hg}$ atom \cite{griffith09}, and the electron (ThO) \cite{ThO14} provide some of the most stringent limits for these beyond-SM $CP$-violating interactions \cite{ermvk13,cm14}. 

Searches in diamagnetic atoms ($^{199}$Hg \cite{griffith09}, $^{129}$Xe \cite{rosenberry01}) and molecules (TlF \cite{cho91}) are primarily sensitive to $CP$-violating interactions in the nucleus \cite{pr05}. Although the nuclear EDM is shielded by the atomic electrons, this shielding is imperfect due to the finite size of the nucleus \cite{schiff}. The remaining measurable quantity, characterized by a radially-weighted EDM of the nucleus, is called the nuclear Schiff moment, which induces a proportional and measurable atomic EDM. The best limits on $CP$-violating interactions originating within the nucleus are derived from the limit on the atomic EDM of $^{199}$Hg: $|$$d$($^{199}$Hg)$|$$<$ 3.1 $\times$ 10$^{-29}$ $e \cdot$cm \cite{swallows13}. Here, we report the first measurement of the EDM of the diamagnetic $^{225}$Ra atom. $P$-odd and $T$-odd effects in $^{225}$Ra are greatly enhanced by collective effects and a closely spaced parity doublet structure resulting from the octupole deformation of its nucleus \cite{PhysRevLett.76.4316}. Indeed, nuclear structure calculations    \cite{PhysRevC.56.1357, PhysRevLett.94.232502}, combined with atomic structure calculations \cite{PhysRevA.66.012111}, predict the EDM of $^{225}$Ra to be 2-3 orders of magnitude larger than that of $^{199}$Hg \cite{PhysRevC.82.015501}. This enhancement factor may be even larger given the sizeable variations in the calculations of $^{199}$Hg \cite{PhysRevC.82.015501}. Meanwhile, these special nuclear structure properties also allow the Schiff moment of $^{225}$Ra to be more reliably calculated than that of $^{199}$Hg \cite{ermvk13}. 

\begin{figure}[t]
\includegraphics[scale=1.1]{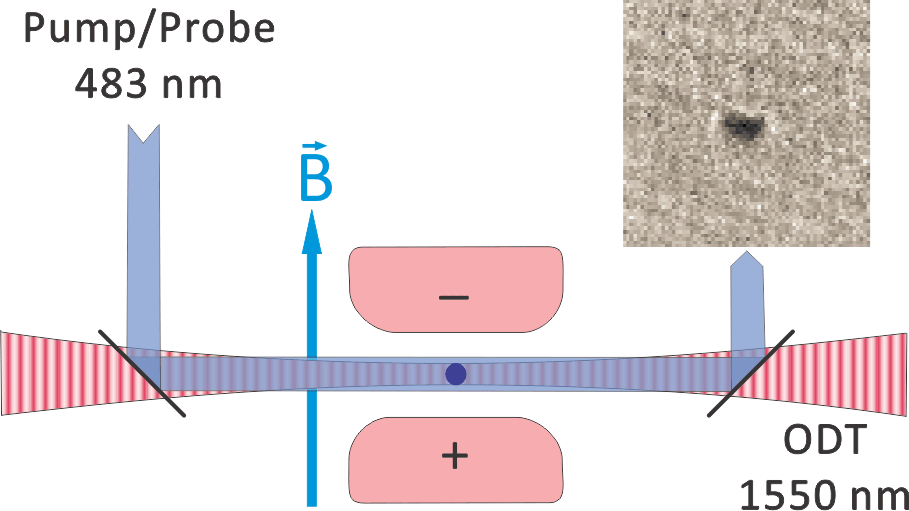}
\caption{(Color online) Diagram of the center of the science chamber. A standing-wave ODT is aligned between a pair of copper electrodes. A collimated beam at 483 nm both optically pumps and probes the atoms, providing absorption images of the atom cloud. A traveling-wave ODT propagating in the direction normal to the page delivers atoms from the 3D MOT to the standing-wave ODT. Inset: CCD image of atoms loaded in the standing-wave ODT; the image is 450 $\mu$m by 450 $\mu$m.}
\label{fig:Diagram}
\end{figure}

$^{225}$Ra is also attractive from an experimental perspective.  The 14.9 day half-life allows this isotope to be obtained as a radioactive source in sufficient quantities for experiments to run off-line, away from an accelerator.  Its nuclear ground state has spin $I=1/2$, which eliminates some systematic effects and decoherence arising from electric field gradients.  Finally, the atomic structure of radium enables laser cooling and trapping \cite{Reference1,PhysRevC.86.065503}. To search for an EDM, spin precession frequencies ($\omega_{\pm}$) under the influence of both a uniform electric field (E) and a uniform magnetic field (B) are measured: $\hbar\omega_{\pm} = 2 \mu B \, \pm \, 2dE $. Here $d$ is the EDM; $\mu$ is the magnetic dipole moment ($-0.7338(15)\mu_{N}$ \cite{PhysRevLett.59.771}); $\omega_{\pm}$ are precession frequencies corresponding to the E-field parallel or anti-parallel to the B-field.  The EDM measurements were performed with $^{225}$Ra atoms in a standing-wave optical dipole trap (ODT) (Fig.\ref{fig:Diagram}).  This approach was theoretically examined in detail for  $^{199}$Hg \cite{PhysRevA.59.4547}. 

$^{225}$Ra is a decay product of the long-lived $^{229}$Th isotope  ($t$$_{1/2}$ = 7300 yr), and can be chemically separated from a $^{229}$Th stock sample.  The U.S. National Isotope Development Center can provide 10-12 mCi of $^{225}$Ra every two months \cite{Shelton}.  For the two experimental runs presented in this paper, $^{225}$Ra samples of 3 mCi and 6 mCi (10$^{14}$ atoms), respectively, were loaded into an oven, and used gradually over the following two weeks.  In addition, the long-lived, naturally-occuring $^{226}$Ra ($t$$_{1/2}$ = 1600 yr) was a useful proxy during the development phase of the experiment, and for tuning the apparatus prior to the EDM measurements.  A typical load of $^{226}$Ra is 2-5 $\mu$Ci, amounting to 10$^{16}$ atoms, or 100 times more atoms than the typical $^{225}$Ra sample.  However, with $I$ = 0, $^{226}$Ra cannot be used in a nuclear spin precession measurement. 


Laser manipulation of radium atoms has previously been described in detail \cite{PhysRevC.86.065503}.  Neutral radium atoms, both $^{225}$Ra and $^{226}$Ra, pass through a transverse cooling region and a Zeeman slower, before being captured in a three-dimensional (3D) magneto-optical trap (MOT) based on the intercombination transition 7s$^{2}$ $^{1}$S$_{0}$ F = $1/2$ $\rightarrow$ 7s7p $^{3}$P$_{1}$ F = $3/2$ at 714 nm. With an associated lifetime of 422 $\pm$ 20 ns \cite{PhysRevA.73.010501}, this transition is weak compared to those typically used for primary cooling and trapping of atoms. However, it is the transition in radium that is the nearest to being closed, with a branching ratio of only 4$\times$10$^{-5}$ leaking to 7s6d $^{3}$D$_{1}$ \cite{Bieron,0953-4075-40-1-021}. Atoms in 7s6d $^{3}$D$_{1}$ are re-pumped to the ground state with a 1429 nm laser.  Largely due to the weakness of the laser force, the trapping efficiency, from the sample in the oven to the MOT, is only 1$\times$10$^{-6}$.  Typically, 10$^{5}$ $^{226}$Ra atoms or 10$^{3}$ $^{225}$Ra atoms are accumulated over the MOT lifetime of 40 seconds and cooled to 40~$\pm$ 15~$\mu$K.  

Next, the atoms are transferred to an optical dipole trap (ODT) formed by focusing a 40 W, 1550 nm, horizontal laser beam with a 2 m focal length lens.  The resulting trap is elongated: 1 cm long and 100 $\mu$m across, with a depth of 500 $\mu$K.  The transfer efficiency from the MOT to the ODT is typically 80\%, helped by the fact that 1550 nm is nearly magic to the cooling transition, meaning the differential light shift of the ground and excited states is comparable to the natural linewidth \cite{0953-4075-40-1-021}.  The ODT is then translated axially over a distance of 1 m into a separate chamber. Here it overlaps a perpendicular standing-wave ODT formed by a single-mode, single-frequency, 10 W, 1550 nm retroreflected laser beam focused down to 100 $\mu$m in diameter, and linearly polarized in the horizontal direction. The transfer of atoms from the first ODT to the second is assisted by briefly turning on a one-dimensional (1D) MOT to compress the aspect ratio of the atom cloud from 100:1 to 1:1.  This 60 $\mu$m cloud retains its shape in the standing-wave ODT after the first ODT is turned off. The overall efficiency of transferring atoms from the 3D MOT to the standing-wave ODT is 5\%. No residual magnetization of the shields (see below) was detected after the 1D MOT was switched off; any residual  magnetic field was found to be less than the measurement sensitivity of 300 nG. 

Figure \ref{fig:Diagram} provides the layout surrounding the atoms in the standing-wave ODT.  The trap is placed at the center of a pair of parallel copper electrodes, cylindrical with a vertical axis, whose end faces are 1.6 cm in diameter and 2.3(1) mm apart. The upper electrode is grounded; the lower one can be ramped to voltages between $+$15.5 kV and $-$15.5 kV, generating a uniform E-field of 67 kV/cm in either the up or down direction (parallel or anti-parallel to the B-field, respectively). The leakage current measured on the grounded side is typically $<$ 80 pA. The variation of the E-field near the center is $<$1\%/mm.  The electrode assembly is inside a glass, nonmagnetic vacuum enclosure, which in turn is surrounded by a cosine-theta coil wound on an aluminum cylinder of 0.32 m diameter and 0.65 m length.  Three layers of mu-metal shields surround this assembly and reduce the influence of external B-fields by a factor of 2 $\times$ 10$^{4}$ when measured in the vertical direction at the center.  The coil inside the shields generates a stable and uniform B-field of 15-30 mG in the vertical direction (intentionally varied between experimental runs).  Its spatial variation is \textless 1\%/cm; its instability is $<$0.01\% when averaged over a load-measurement cycle of 50 s.

The number of atoms in the standing-wave ODT is probed by a blue laser beam tuned to the resonance of 7s$^{2}$ $^{1}$S$_{0}$ F = $1/2$ $\rightarrow$ 7s7p $^{1}$P$_{1}$ F = $3/2$ at 483 nm, co-propagating along the standing-wave ODT laser beams. This transition can cycle for an average of about 1000 times before leaking to lower D states.  The shadow image produced by resonant absorption is cast onto a CCD camera (Fig. \ref{fig:Diagram}).  To produce an absorption image (given about 200 $^{225}$Ra atoms in the trap), the blue beam is pulsed for 1.45 ms, during which each atom absorbs on average 100 photons. A linear combination of images without atoms is used for background subtraction; $i.e.$, to suppress distortions arising from interference effects \cite{PhysRevA.82.061606}.  The noise of this detection scheme is approximately 1.2 times the photon number shot noise.  This detection method is destructive in that it heats the atoms out of the trap and pumps them to metastable levels.  By measuring the number of atoms at various delay times, the lifetime of the atoms in the trap is determined to be 3-5 s, about 10 times shorter than that in the 3D MOT.  This is consistent with the higher vacuum pressure observed in the glass chamber.

The spin polarization is both produced and detected via optical pumping by a circularly polarized blue beam tuned to the resonance of the non-cycling transition 7s$^{2}$ $^{1}$S$_{0}$ F = $1/2$ $\rightarrow$ 7s7p $^{1}$P$_{1}$ F = $1/2$.  An atom with a spin in the fully polarized state does not absorb photons, while one in the opposite-spin state absorbs on average three photons before becoming fully polarized. Compared to the atom-number measurement on a cycling transition, the spin-sensitive detection scatters 30 times fewer photons, resulting in a reduction of image contrast.  

\begin{figure}[t]
\includegraphics[scale=0.3]{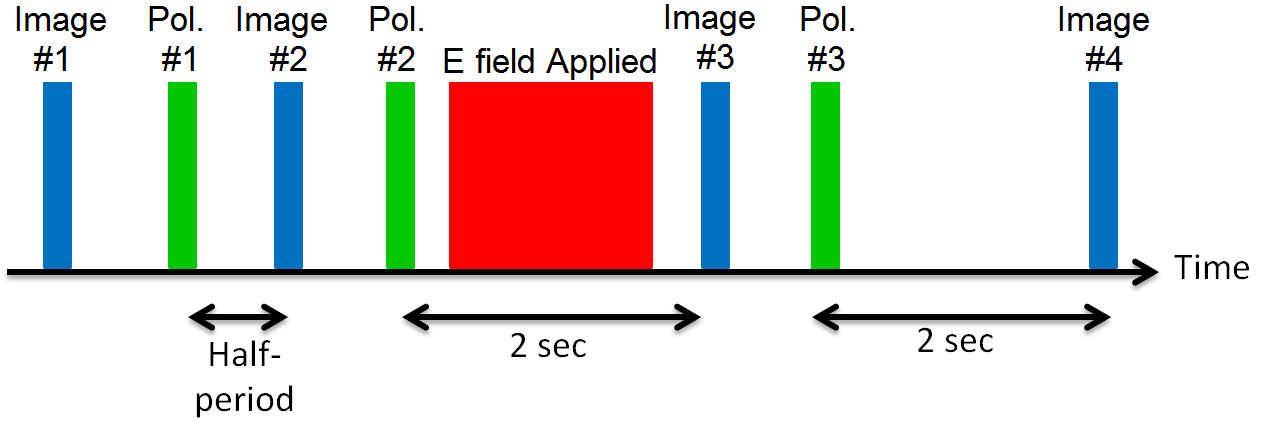}
\caption{(Color online) Pulse sequence used in the EDM measurement. Half-period denotes half of the spin precession period.}
\label{fig:Timings}
\end{figure}

The pulse sequence used for the EDM measurement is shown in Fig. \ref{fig:Timings}. Two kinds of pulses, generated by acousto-optical modulator (AOM) switches, are used: 1.5 ms pulses to polarize the atom cloud, and 60 $\mu$s pulses to measure the number of atoms in the opposite-spin state (short pulses optimize the image signal-to-noise ratio, whereas longer ones for optical pumping maximize atom polarization). After each pulse, the 483 nm laser is blocked within 1 ms by an additional mechanical shutter to prevent decoherence induced by light leaking through the AOM while the atoms precess. Four images of atoms are recorded in each measurement cycle. The first occurs prior to any polarization pulse and, thus, produces a signal at half of the maximum contrast. The second occurs after half of the spin precession period, following a polarization pulse, and has a signal at the maximum contrast. The third is taken about 2 s after polarization; it is during this time that the E-field is applied. Then, the atoms are repolarized, followed by the fourth image also occuring about 2 s after polarization, but this time with no applied E-field. The third and fourth images are normalized to the second one in order to reduce sensitivity to atom number fluctuations. The third image is used to build the ``E-field on" spin precession curve, and the fourth  builds the ``E-field off" spin precession curve. The data indicate that the decreasing contrast is consistent with the lifetime of the atoms in the ODT. 

\begin{figure*}[t]
\includegraphics[scale=0.6]{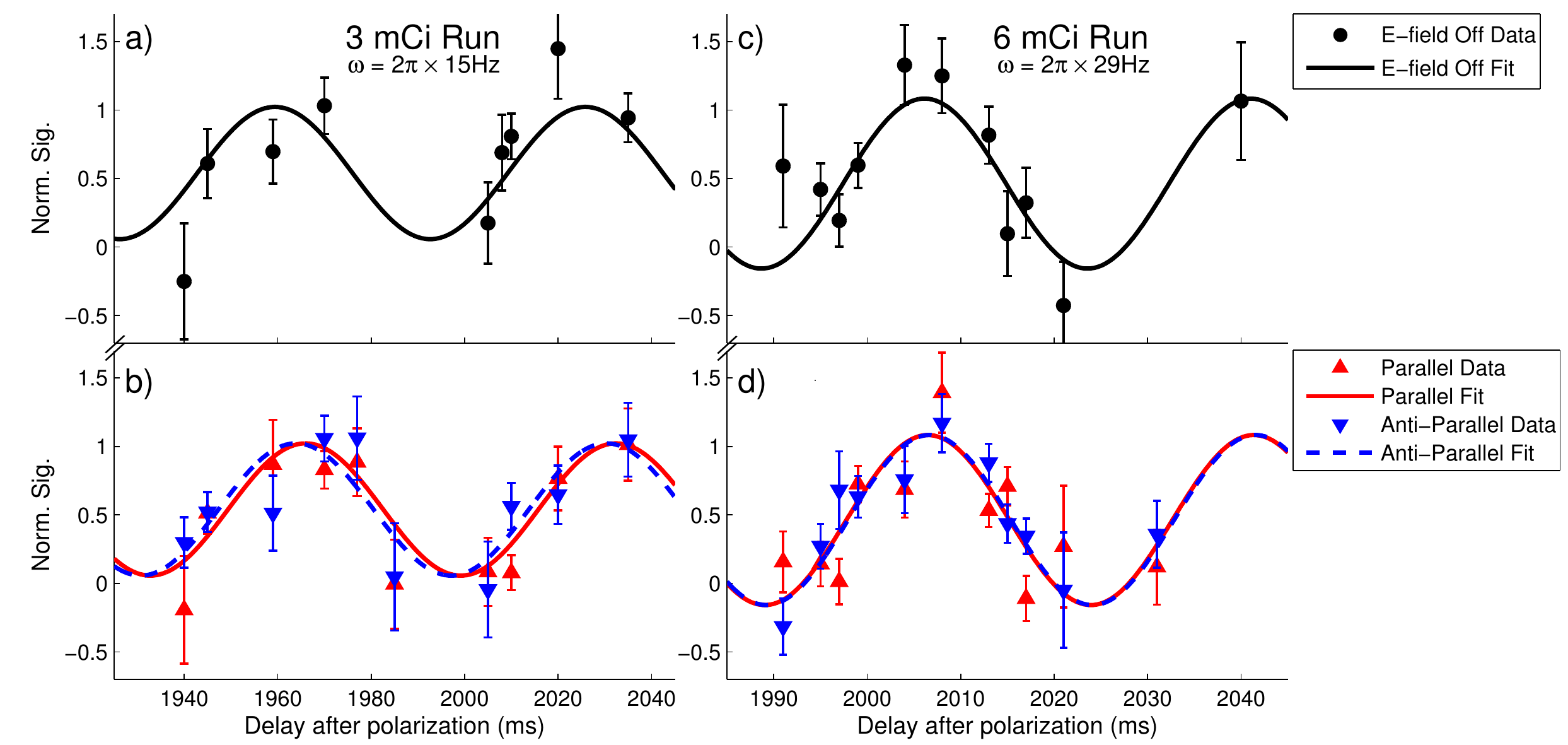}
\caption{(Color online) Precession curves from the two experimental runs. Panels (a) and (b) are based on the first run, which used a 3 mCi $^{225}$Ra source, and panels (c) and (d) are based on the second run with 6 mCi. The panels show data with the E-field parallel to the B-field  (red), E-field anti-parallel to the B-field (blue), and E-field off (black). Between the two runs the bias B-field was deliberately altered, resulting in the two different precession frequencies. An EDM would cause a phase difference between the E-field parallel and E-field anti-parallel fit curves. The global fits for the 3 mCi and 6 mCi runs yield $\chi^2/24 = 1.11$ and $\chi^2/28 = 1.35$, respectively.}
\label{fig:Precession}
\end{figure*}

Figure \ref{fig:Precession} presents the spin detection data with the E-field on,  pointing parallel to the B-field, anti-parallel, or with the E-field off as functions of the free precession time.  The E-field is at its maximum value of 67 kV/cm for 1.2 s during the 2 s of precession. Each data point was corrected for trap losses using the measured trap lifetime. Data taken under the three E-field configurations were simultaneously fit to a combined set of equations: 
\begin{align*}
y_{\mathrm{E-field \, Off}}&= \frac{A}{1+P}\left[1-P\mathrm{cos}(\omega t)\right] \\
y_{\mathrm{Parallel, \, Anti-Parallel}}&= \frac{A}{1+P}\left[1-P\mathrm{cos}(\omega t + \theta \pm \Delta\phi/2 )\right] \\
\end{align*}

Five parameters $A$ (normalization), $P$ (atom polarization), $\omega$ (precession frequency with E-field off), $\theta$ and $\Delta\phi$ (defined below) were fit without constraints. $y_{\mathrm{E-field \, Off}}$ is the integrated signal in Image \#4 after normalization to Image \#2 and background subtraction. Similarly, $y_{\mathrm{Parallel, \, Anti-Parallel}}$ is derived from Image \#3. An EDM would cause a polarity-dependent phase shift, $\Delta\phi$, with the EDM given by $d = \hbar\Delta\phi/(4 E \tau)$ (here $\tau$ is the spin precession time with the E-field applied). An effect common to both E-field polarities would produce an overall phase offset $\theta$. $\Delta\phi$ was found to be uncorrelated with the other fit parameters. In both of the experimental runs, the measured EDM was found to be consistent with zero: $-$$(4.0 \pm 5.2) \times 10^{-22}$ $e \cdot$cm in the first measurement done with 3 mCi of $^{225}$Ra, and $(0.6 \pm 2.9) \times 10^{-22}$ $e \cdot$cm in the second one, done with 6 mCi of $^{225}$Ra. The uncertainties listed above are statistical only. 

A variety of possible systematic effects could potentially lead to false EDM signals. Effects due to correlations between the high voltage and the probe frequency, the external B-field, the current supply for the bias B-field, and the standing-wave ODT power were considered. Also modelled were those due to imperfect E-field reversal, induced B-fields due to the E-field pulsing, leakage current, $\vec{E}$ $\times$ $\vec{v}$ effects, Stark interference in the excited state hyperfine structure due to the presence of the AC laser fields and the DC E-field  \cite{PhysRevA.59.4547}, and geometric phase. The high voltage correlations were measured and, for each systematic effect, models were used to place an upper limit on the size of a potential false EDM. Using this analysis, all the systematic effects are calculated to be smaller than $1 \times 10^{-25}$ $e \cdot$cm. 

The effects due to imperfect E-field switching can also be constrained by direct measurements with trapped $^{225}$Ra atoms. Effects quadratic in E-field do not produce an EDM-like signal, unless the E-field reversal is imperfect. The voltage difference between the electrodes is recorded for each polarity; the E-field imbalance is found to be less than 0.7\%. We use three E-field configurations (parallel, anti-parallel, and off) to directly measure both the linear and quadratic terms; the term that gives the phase a quadratic dependence on E-field allows us to place a limit on any potential $E$$^{2}$ systematics. The ability to place a limit on this effect is, therefore, limited by the statistical uncertainty of the spin precession fit $-$ as this uncertainty improves, the limit placed on this effect will improve as well. For now, the quadratic effect was found to be below 2 $\times$ 10$^{-23}$ $e \cdot$cm for the 3 mCi experimental run and below 5 $\times$ 10$^{-24}$ $e \cdot$cm for the 6 mCi one. 

With the understanding that systematic effects are all negligible at the present level of sensitivity, we measure an EDM $d$$(^{225}$Ra$)$ = $-$(0.5 $\pm$ 2.5$_{\mathrm{stat}}$ $\pm$ 0.2$_{\mathrm{syst}}$) $\times$ 10$^{-22}$ $e \cdot$cm, and, hence, set an upper limit: $|$$d$($^{225}$Ra)$|$ $<$ 5.0 $\times$ 10$^{-22}$ $e \cdot$cm (95\% confidence). This represents the first EDM limit placed on an octupole-deformed species, and the first reported EDM measurement on atoms in an optical dipole trap. 

Improvements on the probing sensitivity by many orders of magnitude are possible.  Detailed analysis indicates that the systematic effects of EDM measurements in an ODT can be reduced to below 10$^{-28}$ $e \cdot$cm \cite{PhysRevA.59.4547}.  Here, we discuss various approaches to improve statistical sensitivity.  A factor of 10 in EDM sensitivity can be gained by increasing the standing-wave ODT lifetime, which can be achieved by improving the vacuum and by reducing heating. The gain in EDM sensitivity is linearly proportional to the gain in lifetime because the load-measurement cycle, currently 50 s, is dominated by the time needed for loading the MOT, and because an EDM measurement can be performed in parallel with MOT loading.

The efficiency of detecting the spin state after the E-field has been turned off can be significantly improved by shelving one spin state to the long-lived $^{3}$D$_{1}$ level.  Stimulated Raman Adiabatic Passage (STIRAP) \cite{RevModPhys.70.1003} can be used to transfer atoms from the 7s$^{2}$ $^{1}$S$_{0}$ F = $1/2$, m$_{\mathrm{F}} = -1/2$  state to 7s6d $^{3}$D$_{1}$, while leaving those in the m$_{\mathrm{F}} = +1/2$ state undisturbed. This step turns the initial task of spin detection into atom-number detection with much higher signal-to-noise ratio, scattering 1000 photons per atom instead of 3. At present, in shadow imaging, the detection noise is limited by the photon noise of the blue probe beam, independent of the atom number. Converting spin detection to atom-number measurement is expected to result in a factor-of-20 gain in EDM sensitivity. 

The trap can be made more efficient by using the 7s$^{2}$ $^{1}$S$_{0}$ F = $1/2$ $\rightarrow$ 7s7p $^{1}$P$_{1}$ F = $3/2$ blue transition for transverse cooling and slowing. Under saturation, the laser force based on this transition is a factor of 100 stronger than the present scheme using the intercombination transition.  With this stronger laser force, trap loading efficiency can be improved by a factor of 30. This scheme, however, requires two additional repump lasers. 

Moreover, stronger sources of $^{225}$Ra are under development at various nuclear physics accelerator facilities, including the Facility for Rare Isotope Beams \cite{doi:10.1142/S0217732314300109}. For example, it has been calculated that spallation of a thorium target induced by a 1 mA beam of deuterons at 1 GeV will yield $^{225}$Ra at the rate of 10$^{13}$ s$^{-1}$ \cite{1306.5024}, which is 5 orders of magnitude stronger than the currently available supply. 

A $^{225}$Ra EDM experiment with an E-field of 100 kV/cm, $N= 1 \times  10^{6}$ atoms, and $\tau = 100$ s of spin precession time can reach a statistical sensitivity at the level of 10$^{-28}$ $e \cdot$cm in $T = 100$ days, according to $ \delta d = \hbar / (2 E \sqrt{\tau N T}) $. Along the way towards this long-term goal, a recent analysis \cite{cm14} suggests that a $^{225}$Ra EDM limit at the level of 10$^{-25}$ $e \cdot$cm would tighten the limits on certain types of $T$- and $P$-violating electron-nucleus interactions by at least an order of magnitude.

We thank J.R. Guest, E.C. Schulte, N.D. Scielzo, I.A. Sulai, and W.L. Trimble for contributions in the earlier development stages, I. Ahmad, H.A. Gould, and D.H. Potterveld for providing advice and assistance with the experiment, and V.A. Dzuba and V.V. Flambaum for providing theoretical guidance. This work is supported by the Department of Energy (DOE), Office of Science, Office of Nuclear Physics, under contract No.'s DEAC02-06CH11357 and DE-FG02-99ER41101. $^{225}$Ra used in this research was supplied by DOE, Office of Science, Isotope Program in the Office of Nuclear Physics. M.N.B., N.D.L., and J.T.S. acknowledge support from Argonne Director's postdoctoral fellowships.

\bibliographystyle{apsrev4-1}
\bibliography{EDM.bbl} 

\end{document}